\documentclass[10pt,journal]{IEEEtran}
\normalsize

\usepackage{cite}
\usepackage{graphicx}
\usepackage[cmex10]{amsmath}
\usepackage{amssymb}
\usepackage{booktabs}
\usepackage{threeparttable}
\usepackage{algpseudocode}
\usepackage{amsfonts}
\usepackage{xcolor}
\usepackage{color}

\usepackage{enumitem}

\usepackage{tabularx} %

\usepackage[font=footnotesize,skip=2pt]{caption}

\usepackage{hyperref}
\hypersetup{pdftex,colorlinks=true,allcolors=black}

\usepackage{dblfloatfix} %

\usepackage{soul}

\usepackage{verbatim}%

\usepackage{makecell}

\usepackage{textcomp}
\def\BibTeX{{\rm B\kern-.05em{\sc i\kern-.025em b}\kern-.08em
		T\kern-.1667em\lower.7ex\hbox{E}\kern-.125emX}}

\newtheorem{lemma}{\bf  Lemma}

\newtheorem{algorithm}{\bf  Algorithm}

\definecolor{green}{RGB}{0,128,0} %

\newcommand{\textrr}[1]{{\color{black} #1}}

\newcommand{\kstar}{k^{\star}}
\newcommand{\mj}{\mathsf{j}}
\newcommand{\AM}{A\&M}

\newcommand{\pade}{Pad\'e}

\newcommand{\PA}{Pad\'e~approximation}

\begin{document}
	\title{Accurate Frequency Estimation with {Fewer} DFT Interpolations based on Pad\'e Approximation}

	\author{
		Kai Wu, 
		J. Andrew Zhang,~\IEEEmembership{Senior Member,~IEEE},
		Xiaojing Huang,~\IEEEmembership{Senior Member,~IEEE}, and \\Y. Jay Guo,~\IEEEmembership{Fellow,~IEEE}
		\vspace{-8mm}
			\thanks{K. Wu, J. A. Zhang, Xiaojing Huang and Y. J. Guo are with the Global Big Data Technologies Centre, University of Technology Sydney, Sydney, NSW 2007, Australia (e-mail: kai.wu@uts.edu.au; andrew.zhang@uts.edu.au;  xiaojing.huang@uts.edu.au; jay.guo@uts.edu.au).}%
	}

\markboth{IEEE Transactions on Vehicular Technology,~Vol.~XX, No.~XX, XXX~2021}
{}
	\maketitle

	\begin{abstract}		
		Frequency estimation is a fundamental problem in many areas. The well-known \AM~and its {variant} estimators have established an estimation framework by iteratively interpolating the discrete Fourier transform (DFT) coefficients. 
		In general, those estimators require two DFT interpolations per iteration, have uneven initial estimation performance against frequencies, and are incompetent for small sample numbers due to low-order approximations involved. 
		Exploiting the iterative estimation framework of \AM, we unprecedentedly introduce the \PA~to frequency estimation, unveil some features about the updating function used for refining the estimation in each iteration, and develop a simple closed-form solution to solving the residual estimation error. Extensive simulation results are provided, validating the superiority of the new estimator over the state-the-art estimators in wide ranges of key parameters.
	\end{abstract}


	\IEEEpeerreviewmaketitle
	
	\vspace{-5mm}

\section{Introduction}\label{sec: intro}

{As a fundamental research issue, frequency estimation of a single-tone complex exponential signal has been investigated in many areas, including vehicular communication and sensing \cite{FreqEst_stallite_wang2019near,thesis_freqEst_2002UTS_Aboutanios,Kai_freqEstComLttr2020}. It is worth noting that vehicular radar/sensing based on communication waveforms is very popular currently \cite{8057284,Andrew_JCASmultiBeamTVT2019}, where estimating target parameters can be treated as frequency estimation problems.}
Discrete Fourier transform (DFT) and its interpolation-based frequency estimation have attracted extensive attention, due to the low complexity and high efficiency. Earlier works exploit the fixed DFT coefficients, which can cause 
uneven estimation bias for different frequencies \cite{thesis_freqEst_2002UTS_Aboutanios}. 
In \cite{FreqEst_TSP2005AandM}, the authors introduced an iterative DFT-interpolated frequency estimator, known as \AM, which improves the asymptotic mean squared error (MSE) performance to $ 1.0147 $ times the Cram\'er-Rao lower bound (CRLB).
Many {variants} of \AM~have been developed, achieving various improvements, e.g., a more accurate initial estimate \cite{Freq_liu2011generalizationAM}, a smaller estimation bias \cite{FreqEst_SP2014AandM_derivatives}, a particular treatment of real sinusoidal signals \cite{FreqEst_SPL2017AandM_derivatives}, and an improved asymptotic performance \cite{FreqEst2019TCOM_Serbes}.

A common issue of the above estimators \cite{FreqEst_TSP2005AandM,Freq_liu2011generalizationAM,FreqEst_SP2014AandM_derivatives,FreqEst_SPL2017AandM_derivatives,FreqEst2019TCOM_Serbes} is that {each iteration requires at least $ 2N $ complex multiplications}, with $ N $ denoting the number of signal samples. This complexity is scaled up with the number of iterations.  
The issue is specifically treated in \cite{Freq_liu2011generalizationAM}
which proposes an improved initial estimate that can reduce up to two interpolations yet with the same asymptotic performance as 
the original \AM. The issue is also noticed in \cite{FreqEst2019TCOM_Serbes} which uses the same number of interpolations as \AM~but achieves a better asymptotic performance.

Another common issue of the estimators \cite{FreqEst_TSP2005AandM,Freq_liu2011generalizationAM,FreqEst_SP2014AandM_derivatives,FreqEst_SPL2017AandM_derivatives,FreqEst2019TCOM_Serbes} is that a first-order Taylor series (TS) is used for approximating the relation between the \textit{updating function} and the estimation error\footnote{As will be illustrated in Section \ref{subsec: core updating function}, the updating function is used for calculating the estimation error in each iteration.}.
On one hand, this makes the estimators incompetent for small $ N $'s, since the first-order TS is only valid for large $ N $'s. On the other hand, the large approximation error can lead to uneven estimation accuracy for different frequencies. As a consequence, more iterations may be required for a satisfactory estimate \cite{FreqEst2019TCOM_Serbes}.  
We underline that the small-$ N $ scenario is not uncommon. The frequency estimator studied here can be used for estimating the angle-of-arrival (AoA) \cite{FreqEst_aoaEstAboutanios2017GRSL,Kai_JCAS_fhMIMO_waveform}, which is known as \textit{spatial frequency}, of an antenna array. The number of antennas available for the spatial DFT is practically limited.

This correspondence is motivated to improve the estimation accuracy of the DFT interpolation-based frequency estimator, reduce interpolations and enhance the applicability to wider range of sample numbers. 
	A key innovation of this work is unprecedentedly introducing the \PA~(PA) to the interpolation-based frequency estimation.
	{PA employs the ratio of two polynomials to approximate a given power series, e.g., TS, where the sum of the degrees of the two polynomials in a PA is equal to the degree of the power series \cite[Sec. 5.12]{book_press2007numerical}.} 
	 Underlying the successful application of PA are several key contributions, as summarized below.

We construct the PA that approximates the sixth-order TS of the updating function (in contrast to the first-order TS employed in most previous designs \cite{FreqEst_TSP2005AandM,Freq_liu2011generalizationAM,FreqEst_SP2014AandM_derivatives,FreqEst_SPL2017AandM_derivatives,FreqEst2019TCOM_Serbes}). We prove some useful features of the updating function in terms of its monotonicity and symmetry, and use the features to simplify the PA by suppressing some high-order terms. 
We also derive the coefficients of the simplified PA.
  Moreover, we develop a closed-form solution to solving the frequency estimation error from the established PA, where \textrr{we} apply the features of the updating function to remove the ambiguities in the solution. 
Extensive simulations validate that our estimator achieves more uniform performance 
across wide ranges of sample numbers and frequencies, and approaches CRLB more tightly, compared with state-of-the-art estimators.

\vspace{-2mm}
 
\section{Signal Model} \label{sec: signal model and QSE}
Consider the estimation of the unknown frequency $ f $ from the following $ N $ samples of a complex single-tone exponential signal,
\begin{align}\label{eq: complex sinusoidal signal}
	s(n) = Ae^{\mj\left( \frac{2\pi fn}{f_s} +\phi\right)} + z(n),~n=0,1,\cdots,N-1,
\end{align}
where $ A $, $ f_s $, $ \phi $ and $ N $ denote the signal amplitude, the sampling rate, the initial phase and the sample number, respectively. The additive white Gaussian noise (AWGN)
is denoted by $ z(n) $. 
The frequency $ f $ can be decomposed into 
the sum of an integer and a fractional multiples of $ f_{\mathrm{s}}/N $, i.e., 
\begin{align} \label{eq: f decomposition}
	f={(k^{\star}+\delta)}f_s\Big/N,~\delta\in [-0.5,0.5],
\end{align}
where $ k^{\star} $ denotes an integer and $ \delta $ a non-integer. Taking the $ N $-point DFT of $ s(n) $, i.e., $ \sum_{n=0}^{N-1}s(n)e^{-\mj\frac{2\pi kn}{N}} $, yields, 
\begin{align} \label{eq: S(k)=sum... DFT}
	S(k) = \sum_{n=0}^{N-1}\tilde{A}e^{\mj \frac{2\pi (\kstar+\delta)n}{N}}e^{-\mj\frac{2\pi kn}{N}} + Z(k),
\end{align}
where $ \tilde{A}=Ae^{\mj \phi} $, $ f $ in (\ref{eq: complex sinusoidal signal}) is replaced by its expression given in (\ref{eq: f decomposition}), and $ Z(k) $ denotes the DFT of $ z(n) $. 
{The integer $ \kstar $ can be estimated by identifying the maximum of $ |S(k)|^2 $, i.e., 
	\begin{align}\label{eq: kstar}
		\kstar:~\max{}_{k\in[0,N-1]}|S(k)|^2.
\end{align}}
\vspace{-4mm}

As generally treated in previous works \cite{FreqEst_TSP2005AandM,Freq_liu2011generalizationAM,FreqEst_SP2014AandM_derivatives,FreqEst_SPL2017AandM_derivatives,FreqEst2019TCOM_Serbes}, we assume that {$ \kstar $ is accurately estimated so as to {focus on refining the frequency estimate by estimating $ \delta $}}. 
This is a legitimate assumption, since we tend to focus on the asymptotic performance of a frequency estimator, as achieved in high SNR regions, when developing a new estimator or comparing with the previous ones.
Also following the related works \cite{FreqEst_TSP2005AandM,Freq_liu2011generalizationAM,FreqEst_SP2014AandM_derivatives,FreqEst_SPL2017AandM_derivatives,FreqEst2019TCOM_Serbes}, we focus on the single-tone signal to introduce the new ideas/designs. The extension to multi-tone scenarios will be remarked in Section \ref{sec: conclusion}.

\section{Proposed Frequency Estimator} \label{sec: proposed estimator}

In this section, a new estimator is developed to iteratively estimate $ \delta $. We first introduce the core updating function, based on which we establish the overall estimator. 

\vspace{-3mm}

\subsection{Core Updating Function} \label{subsec: core updating function}
Consider the $ i $-th iteration, where the estimate of $ \delta $ from the previous iteration, as denoted by $ \delta_{i-1} $, is available. With reference to \cite{FreqEst2019TCOM_Serbes}, we interpolate the DFT coefficients at 
\begin{align} \label{eq: k_{i,+-}}
	k_{i,\pm}=\kstar + \delta_{i-1} \pm q_i,
\end{align}
where $ \kstar $ is given in (\ref{eq: f decomposition}), and $ q_i $ is an extra controlling parameter used to shift the interpolation positions.
{Plugging $ k=k_{i,\pm} $ into (\ref{eq: S(k)=sum... DFT}), the interpolated DFT coefficients are
	\begin{align}\label{eq: S(k_i,+-)}
		S(k_{i,\pm}) = \breve{A}{\sin(\pi(\xi_i\mp q_i))}\big/{\sin({\pi(\xi_i\mp q_i)}/{N})},
	\end{align}
	where $ \breve{A}=\tilde{A}e^{\mj\frac{(N-1)\pi}{N}(\xi_i\mp q_i)} $, $ \tilde{A} $ given in (\ref{eq: S(k)=sum... DFT}), $ \xi_i(=\delta-\delta_{i-1}) $ denotes the estimation error in the $ (i-1) $-th iteration, and the noise term is dropped for brevity.} Denoting $ S_{i,\pm} = S(k_{i,\pm}) $, we can construct the following ratio,
\begin{align}\label{eq: rho_i}
	\rho_i ={(|S_{i,+}|^2-|S_{i,-}|^2)}\Big/{(|S_{i,+}|^2+|S_{i,-}|^2)}
\end{align}
The purpose of doing so is to estimate $ \xi_i $ from $ \rho_i $. Let 
$ \hat{\xi}_i $ denote the estimate of $ \xi_i $. We can use $ \hat{\xi}_i $ to refine the $ \delta $ estimation as follows \cite{FreqEst2019TCOM_Serbes,FreqEst_TSP2005AandM},
\begin{align} \label{eq: delta_i=delta_{i-1}+...}
	\delta_i=\delta_{i-1} + \hat{\xi}_i,~\mathrm{s.t.}~\hat{\xi}_i\approx\xi_i=\delta-\delta_{i-1}.
\end{align}
Next, we illustrate how to estimate $ \hat{\xi}_i $ from $ \rho_i $.

Introducing the function
$ \mathcal{S}(\xi_i,\pm q_i) =  \frac{\sin^2(\pi(\xi_i\mp q_i))}{\sin^2(\frac{\pi(\xi_i\mp q_i)}{N})} $,
the right-hand side (RHS) of (\ref{eq: rho_i}) can be written into
\begin{align} \label{eq: f( xi_1 )}
	f(\xi_i) = \frac{ \mathcal{S}(\xi_i,q_i) - \mathcal{S}(\xi_i,-q_i) }{ \mathcal{S}(\xi_i,q_i) + \mathcal{S}(\xi_i,-q_i) }.
\end{align}	
Jointly observing (\ref{eq: rho_i}) and (\ref{eq: f( xi_1 )}), we see that estimating $ \hat{\xi}_i $ from $ \rho_i $ is equivalent to solving the equation $ f(\xi_i)=\rho_i $. 
Some features of $ f(\xi_i) $, which are useful for \textrr{deriving the solution}, are provided in the following lemma; refer to Appendix \ref{app: proof monotonicity and symmetry} for its proof.

\vspace{2mm}

\begin{lemma} \label{lm: rho monotonic}
	\textit{For $ |\xi_i|\le q_i $, $ f(\xi_i) $ monotonically increases with $ \xi_i $, and {presents odd symmetry around the origin, i.e., $ f(-\xi_i) = -f(\xi_i) $.}}
\end{lemma}

\vspace{2mm}

From (\ref{eq: f( xi_1 )}), we see that it can be difficult to directly solve the equation $ f(\xi_i)=\rho_i $, due to the existence of the squared sine functions. To simplify the equation, many previous estimators, e.g., \cite{FreqEst_TSP2005AandM,Freq_liu2011generalizationAM,FreqEst_SP2014AandM_derivatives,FreqEst_SPL2017AandM_derivatives,FreqEst2019TCOM_Serbes,FreqEst_quinn1994estimating}, use the first-order TS of $ f(\xi_i) $ which is valid for large $ N $'s. In contrast, we propose to use the PA to approximate the following TS of $ f(\xi_i) $ (of degree six), i.e.,
\begin{align} \label{eq: tilde f(xi_i) TS}
	\tilde{f}(\xi_i)=\sum_{l=0}^{6}c_l\xi_i^l=c_1\xi_i + c_3\xi_i^3 + c_5\xi_i^5, 
\end{align}
where the even powers of $ \xi_i $ are suppressed since $ f(\xi_i) $ is an odd function of $ \xi_i $, as illustrated in Lemma \ref{lm: rho monotonic}. 
We propose to use the following PA to approximate the above TS,
\begin{align}\label{eq: hat f(xi_i) PA}
	& \hat{f}(\xi_i)={\left(\sum_{p=0}^{P} a_p \xi_i^p\right) }\left/{ \left(\sum_{r=0}^{R} b_r \xi_i^r\right) }\right.,~P=R=3, \\
	&\mathrm{s.t.}~ \hat{f}(0)=\tilde{f}(0);~\hat{f}^{(k)}(0)=\tilde{f}^{(k)}(0),~k=1,\cdots,P+R,\nonumber
\end{align}
where $ {h}^{(k)}(x) $ denotes the $ k $-th derivative of $ h(x) $ ($ h $ can be $ \hat{f} $ or $ \tilde{f} $) and $ {h}^{(k)}(0) $ is value of $ {h}^{(k)}(x) $ at $ x=0 $. 
Note that {the rationale for setting $ P=R=3 $ is illustrated in Appendix \ref{app: rationale of P=R=3}}. 

Also note that the constraints in (\ref{eq: hat f(xi_i) PA}) constitute $ (P+R+1) $ equations, which can be used to express the coefficients of the PA, i.e., $ a_p $ and $ b_r $, in terms of those of TS, i.e., $ c_l $. 
%
We underline that the properties of $ f(\xi_i) $ unveiled in Lemma \ref{lm: rho monotonic} can be used to suppress some high-order terms in $ \hat{f}(\xi_i) $. 
As illustrated in Appendix \ref{app: rationale of P=R=3}, we have $ a_0=a_2=0 $ and $ b_1=b_3=0 $. 
Plugging these constraints into the $ (P+R+1) $ equations, the PA coefficients can be solved, with the following solution achieved.

\vspace{2mm}

\begin{lemma} \label{lm: pade approximation of rho}
	{\it
		The function $ f(\xi_i) $ can be approximated by $ \hat{f}(\xi_i) $ with the approximation error in the order of $ \mathcal{O}(\xi_i^7) $, where 
		\begin{align}
			& \hat{f}(\xi_i) = {(a_1 \xi_i + a_3 \xi_i^3)}\Big/{(1 + b_2 \xi_i^2)},\nonumber\\
			\mathrm{with}~& a_1 = c_1,a_3=c_3-{c_1c_5}\big/{c_3},b_2={c_5}\big/{c_3}.
		\end{align}
	}
\end{lemma}

\vspace{2mm}

Equating $ \hat{f}(\xi_i) $ obtained in Lemma \ref{lm: pade approximation of rho} to $ \rho_i $ calculated in (\ref{eq: rho_i}), we obtain that $ a_1 \xi_i + a_3 \xi_i^3 = \rho(1 + b_2 \xi_i^2), $
which can be further turned into a cubic equation of $ \xi_i $,
\begin{align} \label{eq: cubic equatin}
	& \xi_i^3 +k_2 \xi_i^2 + k_1 \xi_i + k_0	= 0,\nonumber\\
	\mathrm{s.t.}~& k_2=-{\rho b_2}/{a_3},k_1={a_1}/{a_3},k_0=- {\rho}/{a_3},
\end{align}
Using the cubic formula \cite{web_cubicFormulaWolfram}, the three roots of the above equation can be given by
\begin{align}\label{eq: roots of cubic equation}
	&~z_1 = -k_2/3 + 2B, ~z_2 = -k_2/3 - B + D, \nonumber\\
	&~z_3 = -k_2/3 - B - D,
\end{align}
where $ B=(S+T)/2 $ and $ D={\sqrt{3}}(S-T)\mj/2 $; 
$ S=\sqrt[3]{R+\sqrt{D}}$ and $T=\sqrt[3]{R-\sqrt{D}}$; $D=Q^3 + R^2, R={(9k_1k_2-27k_0-2k_2^3)}/{54}$ and $  Q={(3k_1-k_2^2)}/{9}$;
and $ k_0,k_1 $ and $ k_2 $ are given in (\ref{eq: cubic equatin}). 
To determine the final estimate of $ \xi_i $, we provide the following.

\vspace{2mm}
\begin{lemma}\label{lm: estimate of xi_i}
	{\it 
	The estimate of $ \xi_i $ is given by 
	\begin{align} \label{eq: xi i hat=smallest root}
		\hat{\xi}_i=z_{i^*},~\mathrm{s.t.}~i^*=\mathrm{argmin}_i~|z_i|.
	\end{align}
}
\end{lemma}

\begin{IEEEproof}
	As proved in Lemma \ref{lm: rho monotonic}, 
	$ f(\xi_i) $ is monotonic against $ \xi_i $ for $ |\xi_i|<q_i $. Thus, we can only have one solution to $ f(\xi_i)=\rho_i $ in the region of $ |\xi_i|<q_i $. Since the continuous region covers $ \xi_i=0 $ (the smallest value that can be taken), the solution to the equation $ \hat{f}(\xi_i)=\rho_i $ in the region is \textrr{the smallest root given in (\ref{eq: roots of cubic equation})}. This leads to the (\ref{eq: xi i hat=smallest root}). 
\end{IEEEproof}

Based on the above analyses and derivations, we summarize below the steps of estimating $ \delta_i $ from $ \delta_{i-1} $.

\vspace{2mm}
\begin{algorithm} \label{alg: xi_i estimation}
	{\it
	Given $ a_1,a_3 $, $ b_2 $, $ \delta_{i-1} $ from iteration $ (i-1) $ and $ q_i $, and provided that $ |\xi_i|\le q_i $, $ \delta_i $ can be estimated as follows: 
	\begin{enumerate}\normalfont
		\item Interpolate the DFT coefficients at $ k_{i,\pm} $ given in (\ref{eq: k_{i,+-}}), leading to $ S_{i,\pm} $ given in (\ref{eq: rho_i}); \label{stepA1: interpolation}
		
		\item Construct $ \rho_i $, as illustrated in (\ref{eq: rho_i});
		
		\item  Compute the coefficients $ k_0,k_1 $ and $ k_2 $ based on (\ref{eq: cubic equatin});
		
		\item Compute the three roots in (\ref{eq: roots of cubic equation});
		
		\item Obtain the estimate of $ \xi_i $, as given in Lemma \ref{lm: estimate of xi_i};
		
		\item Update $ \delta_{i} $ as done in (\ref{eq: delta_i=delta_{i-1}+...}). 
	\end{enumerate}
}
\end{algorithm}

\subsection{Initialization and Proposed Estimator in Overall}

As revealed in \cite{FreqEst2019TCOM_Serbes}, \textrr{a high-quality initialization} can speed up the asymptotic convergence of an iterative frequency estimator. Below, we provide a high-quality initialization of the proposed estimator with a single interpolation; c.f., two or more interpolations \textrr{in many previous designs} \cite{FreqEst_TSP2005AandM,Freq_liu2011generalizationAM,FreqEst_SP2014AandM_derivatives,FreqEst_SPL2017AandM_derivatives,FreqEst2019TCOM_Serbes,FreqEst_quinn1994estimating}. 

Assume that $ \delta\in [0,0.5] $ holds. 
If we set $ \delta_0=0.25 $, then the estimation error $ \xi_1 $ satisfies that $ \xi_1=\delta-\delta_0 \in [-0.25,0.25]. $
Accordingly, we can set $ q_1=0.25 $ and run Algorithm \ref{alg: xi_i estimation} to estimate $ \delta_1 $. Moreover, we notice that $ \delta_0-q_1=0 $. This indicates that one of the interpolated DFT coefficients is at $ k_{1,-}=\kstar+\delta_0-q_1=\kstar $. This DFT coefficient has been computed when identifying $ \kstar $; \textrr{see (\ref{eq: kstar}) in Section \ref{sec: signal model and QSE}}. Thus, we only need to interpolate one DFT coefficient at $ k_{1,+} = \kstar + \delta_0 + q_1=0.5 $. 
The above analysis also applies for the case of $ \delta\in [-0.5,0] $. 
\textrr{Then, the next question is how to determine the initial region of $ \delta $.} To do so, the following sign test \cite{Freq_liu2011generalizationAM} can be performed by reusing the DFT results for identifying $ \kstar $ ,
\begin{align} \label{eq: alpha sign test}
	\alpha=\mathrm{sign}\{ [S({k}^{\star}-1)-S({k}^{\star}+1)] S^{\dagger}({k}^{\star})\}, 
\end{align}
where $ \kstar $ is given in (\ref{eq: f decomposition}), and $ ()^{\dagger} $ takes the complex conjugate. Using $ \alpha $, we have 
\begin{align} \label{eq: delta>0 alpha>0, delta<0 alpha <0}
	\delta\in [0,0.5],~\mathrm{if}~\alpha >0;~\mathrm{or~}\delta\in	[-0.5,0],~\mathrm{if}~\alpha < 0.
\end{align}
{As will be illustrated in Section \ref{sec: simulations}, the sign test has 
a high accuracy in the sense that {the estimators with or without using the sign test approach the CRLB from the same SNR}.}

Based on the above initialization and Algorithm \ref{alg: xi_i estimation}, we summarize the overall processing of the proposed frequency estimator in the following algorithm. 

\vspace{2mm}

\begin{algorithm} \label{alg: overall estimator}
	\textit{Input: $ N $, $ \delta_0 = 0.25\alpha $, $ q_1 = 0.25 $, $ q_i~(i=2,\cdots,I) $, the coefficient set $ \mathcal{C}_1 = \{a_1,a_3, b_2\} $ for $ \delta_1 $, and the sets $ \mathcal{C}_i = \{a_1^{(i)},a_3^{(i)}, b_2^{(i)}\} $ for $ \delta_i~(i\ge 2) $. The proposed estimator performs as follows:}
	\begin{enumerate}		
		\item Estimate $ \delta_1 $ by running Algorithm \ref{alg: xi_i estimation} once based on $ N $, $ \delta_0 $, $ q_1 $ and $ \mathcal{C}_1 $;
		
		\item For each $ i=2,\cdots,I $, run Algorithm \ref{alg: xi_i estimation} iteratively based on $ N $, $ \delta_{i-1} $, $ q_i $ and $ \mathcal{C}_i $.
	\end{enumerate}	
	\textit{The final frequency estimate is given by $ \hat{f}=\frac{f_{\mathrm{s}}(\kstar+\hat{\delta}_I)}{N} $.}
\end{algorithm}

\vspace{2mm}

We remark that, it is non-trial to analyze the (asymptotic) performance of the proposed estimator. 
The main reason is because the analysis strategy, as developed in \cite{FreqEst_quinn1994estimating} and widely used in previous works \cite{FreqEst_TSP2005AandM,Freq_liu2011generalizationAM,FreqEst_SP2014AandM_derivatives,FreqEst_SPL2017AandM_derivatives,FreqEst2019TCOM_Serbes}, relies on a linear approximation between $ \rho_i $ and $ \xi_i $; whereas, in contrast, we use the non-linear PA to pursue a high-accuracy depiction of the relation between the two variables. Nevertheless, through extensive simulations to be provided in Section \ref{sec: simulations}, our estimator is seen to outperform several state-of-the-art estimators which have the theoretical guarantee of approaching the CRLB. Two other remarks on Algorithm \ref{alg: overall estimator} are provided below.

\subsubsection{Computational Complexity (CC)}
{To analyze the overall CC of the proposed estimator, we first evaluate the CC of Algorithm \ref{alg: xi_i estimation}. \textrr{Its} CC is dominated by that of interpolating DFT coefficients in Step \ref{stepA1: interpolation}). According to (\ref{eq: S(k)=sum... DFT}), \textrr{one interpolation}
	needs $ N $ complex multiplications (CMs). As illustrated above (\ref{eq: alpha sign test}), a single interpolation is required for the first iteration, while for iteration $ i(\ge 2) $, we need $ 2N $ CMs to interpolate the DFT coefficients twice. Algorithm \ref{alg: overall estimator} runs Algorithm \ref{alg: xi_i estimation} for $ I $ times, and thus its overall CC is dominated by $ N+(I-1)\cdot 2N=(2I-1)N $ CMs.
	In contrast, the CC of the previous estimators, e.g., the three benchmarks to be illustrated in Section \ref{subsec: benchmark estimators for simulation}, are dominated by $ 2IN $ CMs.}  
As will be illustrated in Section \ref{sec: simulations}, 
\textrr{the proposed estimator is able to approach the CRLB after only two iterations, i.e., $ I=2 $.}

\subsubsection{Selection of $ q_i $}
We recommend taking $ q_i=q_1=0.25~(\forall i) $ given two reasons. \textit{First}, the proposed estimator is robust against the value of $ q_i $, or in other words the estimation performance remains almost the same across a wide range of $ q_i $'s. This will be illustrated in Fig. \ref{fig: mse vs q} of Section \ref{sec: simulations}. \textit{Second}, a benefit of taking of $ q_i=q_1=0.25~(\forall i) $ is that only \textrr{one set of PA coefficients, i.e., $ \mathcal{C}_1 $, are required to be stored} onboard, saving storage and time for indexing different sets (if used) in practical systems.

\section{Simulation Results}\label{sec: simulations}
In this section, we provide simulation results to validate the superior performance of the proposed estimator over the previous estimators.

\subsection{Benchmark Estimators}\label{subsec: benchmark estimators for simulation}

Several related estimators are simulated as benchmarks
which can be implemented in the framework of Algorithm \ref{alg: overall estimator}. Below, we only highlight their differences from our estimator.

\textit{1) \AM \cite{FreqEst_TSP2005AandM}:}
This estimator always interpolates the DFT coefficients at $ k_{i,\pm}=\kstar+\delta_{i-1}\pm0.5~(\forall i\ge 1) $ with $ \delta_0=0 $ taken. The ratio $ \rho_i =  \frac{|S(k_{i,+})|-|S(k_{i,-})|}{|S(k_{i,+})|+|S(k_{i,-})|} $ is constructed in each iteration, and $ \xi_i=\rho_i/2 $. \AM~also has a different construction of $ \rho_i $ which leads to the same asymptotic performance as the one given above and hence is not considered here.

\textit{2) Generalized \AM~(GAM) \cite{Freq_liu2011generalizationAM}:}
This estimator iterates as \AM~but starts from a different initial $ \delta_0 $. In particular, GAM takes $ \delta_0=\alpha\beta $, where $ \alpha $ is given in (\ref{eq: alpha sign test}) and $ \beta=0,0.25 $ and $ 0.5 $ are considered and compared in the work. Here, for a fair comparison with our estimator, we only consider $ \beta=0.25 $.

\textit{3) Hybrid \AM~and $ q $-Shift Estimator (HAQSE) \cite{FreqEst2019TCOM_Serbes}:}
This estimator applies \AM~for $ \delta_1 $. Then, starting from $ i=2 $, HAQSE interpolates the DFT coefficients at $ k_{i,\pm} = \kstar+\delta_{i-1}\pm q_{\mathrm{H}} $, where $ q_{\mathrm{H}}=N^{-1/3} $ is proven to be sufficient for the estimator to converge to the CRLB for large $ N $'s and $ q_{\mathrm{H}}\le 0.32 $ is suggested in \cite{Kai_freqEstComLttr2020} to ensure the validity of HAQSE also for small $ N $'s. HAQSE constructs $ \rho_i $ as $ \rho_i=\Re\left\{ \frac{S(k_{i,+})-S(k_{i,-})}{S(k_{i,+}) + S(k_{i,-})} \right\} $ and updates $ \xi_i= \frac{q_{\mathrm{H}}\cos^2(\pi q_{\mathrm{H}})}{1-\pi q_{\mathrm{H}}\cot(\pi q_{\mathrm{H}})} \rho_i $.

\subsection{Results and Analysis}

Unless otherwise specified, the following parameters are used for all the estimators: $ \kstar=2 $, $ f_{\mathrm{s}}=1 $, $ I=2 $, $ \delta\in\mathcal{U}_{[-0.5,0.5]} $, $ q_1=q_2=0.25 $ (for the proposed estimator), and $ q_{\mathrm{H}}=N^{-1/3} $ (for HAQSE). Note that $ \mathcal{U}_{[-0.5,0.5]} $ denotes the uniform distribution in $ [-0.5,0.5] $. All the results to be presented are averaged over $ 5\times 10^4 $ independent trials. 
Moreover, the CRLB \cite{FreqEst_TSP2005AandM}, given by $ \frac{6f_{\mathrm{s}}^2}{4\pi^2\gamma N^3} $, is provided in most simulation results,
where $ \gamma=\frac{|A|^2}{\sigma_0^2} $ is the SNR of the single-tone signal given in (\ref{eq: complex sinusoidal signal}), and $ \sigma_0^2 $ denotes the noise variance of $ z(n) $ therein. 
{As interpreted in Table \ref{tab: marker definitions}, different estimators in the simulation results are differentiated by markers, while different values of $ N $ are distinguished by line styles.}

\begin{table}[!t]\footnotesize
	\captionof{table}{Marker and Line Style Definitions in Simulation Results.}
	\vspace{-2mm}
	\begin{center}
		\begin{tabular*}{78mm}{c| ccccc}
			\hline
			Marker    & $ \circ $ & $ + $ & $ \times $ & $ \triangledown $ & none \\				
			\hline
			Estimator & AM   & GAM & HAQSE & Proposed & CRLB \\				
			\hline\hline
		\end{tabular*}		
		\begin{tabular*}{50mm}{c| ccc}
			Line style & dash   & dash-dotted & solid \\				
			
			\hline
			$ N $    & $ 8 $ & $ 16 $ & $ 32 $ \\				
			\hline			
		\end{tabular*}
	\end{center}
	\vspace{-4mm}
	\label{tab: marker definitions}
\end{table}

\begin{figure}[!t]
	\centering{\includegraphics[width=85mm]{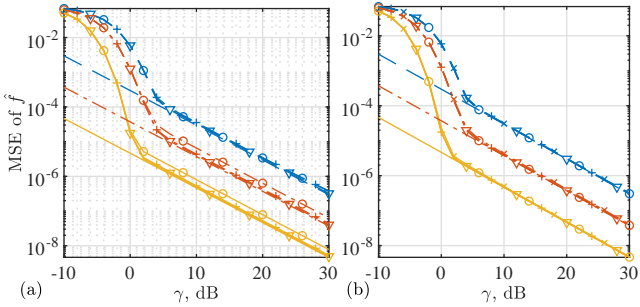}}
	\caption{MSE of $ \hat{f} $ versus $ \gamma $: (a) is for the first iteration, (b) for the second.}
	\label{fig: mse vs snr diff N}
	\vspace{-3mm}
\end{figure}

\begin{figure}[!t]
	\centering{\includegraphics[width=85mm]{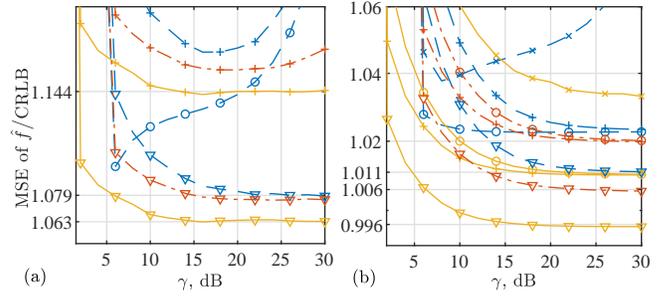}}
	\caption{Illustration of the ratio between MSE and CRLB to better compare different estimators: (a) for the first iteration, (b) for the second. For better clarity, range limits are imposed on the y-axes, and hence the estimators with larger MSEs become invisible in the figure.}
	\label{fig: mse2crb vs snr diff N}
	\vspace{-3mm}
\end{figure}

{Fig. \ref{fig: mse vs snr diff N} plots the MSE of $ \hat{f} $ against $ \gamma $, where $ N=8,16 $ and $ 32 $ are simulated}. 
We see that the MSEs of $ \hat{f} $ converge to the CRLB for all the estimators. 
Comparing the two sub-figures, it is obvious that the estimation performance of all the estimators is further improved (closer to the CRLB) after the second iteration. Fig. \ref{fig: mse2crb vs snr diff N} zooms in the differences among the estimators by normalizing the MSEs plotted in Fig. \ref{fig: mse vs snr diff N} \textrr{against their} respective CRLBs. We see from Fig. \ref{fig: mse2crb vs snr diff N}(a) that, after the first iteration, the proposed estimator
already achieves the MSE as low as $ 1.079 $ times the CRLB for a small $ N=8 $, and reduces the MSE to $ 1.063 $ times the CRLB for $ N=32 $, which is notably based on a single interpolation.
We see from Fig. \ref{fig: mse2crb vs snr diff N}(b) that, after the second iteration, the proposed estimator persistently outperforms the benchmark estimators across the whole region of $ \gamma $ and approaches the CRLB more tightly. 

\begin{figure}[!t]
	\centering{\includegraphics[width=85mm]{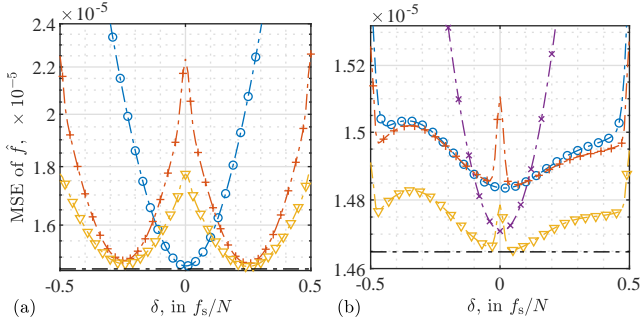}}
	\caption{MSE performance versus $ \delta $, where (a) for the first iteration, and (b) for the second. Note that $ N=16 $, $ \gamma=20 $ dB, and $ 100 $ discrete values are evenly taken in the region of $ \delta\in[-0.5,0.5] $. }
	\label{fig: mse vs delta}
	\vspace{-3mm}
\end{figure}

{We see from Fig. \ref{fig: mse2crb vs snr diff N}(b) that the simulated MSE can be smaller than the CRLB, yet with a considerably small difference.} 
Two reasons may cause this phenomenon. \textit{First}, CRLB is derived for a deterministic parameter that is under estimation, while the frequency taken for the simulations is random over independent trials. 
We remark that this random configuration is necessary for a fair comparison of different estimators, since they can have dramatically distinct estimation performance over frequencies, as will be illustrated in Fig. \ref{fig: mse vs delta}. \textit{Second}, this phenomenon can be caused by the finite number of independent (Monte-Carlo) trials. Refer to \cite[Sec. V]{FreqEst2019TCOM_Serbes} for a detailed analysis of this aspect.

\begin{figure}[!t]
	\centering{\includegraphics[width=85mm]{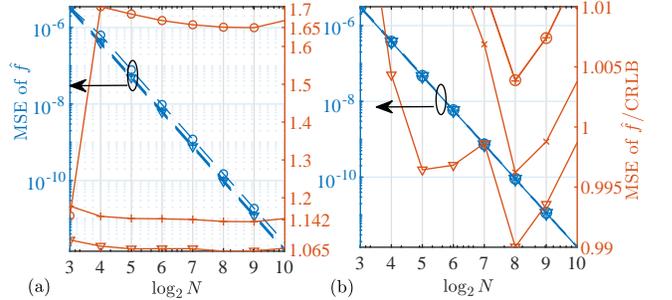}}
	\caption{MSE performance against $ N $, where $ \gamma=20 $ dB, (a) for the first iteration, and (b) for the second.}
	\label{fig: mse vs N}
	\vspace{-3mm}
\end{figure}

Fig. \ref{fig: mse vs N} illustrates the MSE performance w.r.t. $ N $. From Fig. \ref{fig: mse vs N}(a), we see that, after the first iteration, the normalized MSEs of the proposed estimator is as low as $ 1.065 $ which is improved by about $ 6.74\% $ over \AM~and GAM (both achieve the minimum about $ 1.142 $). We also see that the proposed estimator present a more stable asymptotic performance than \AM~and GAM, as $ N $ increases.
From Fig. \ref{fig: mse vs N}(b), we see that, after the second iteration, the proposed estimator is able to approach the CRLB for almost all values of $ N $, while HAQSE can only achieve this for large $ N $'s. This is expected since HAQSE is designed for large $ N $'s. On the other hand, this \textrr{highlights} the benefit of introducing the \pade~approximation.

Fig. \ref{fig: mse vs delta} plots the MSE of different estimators against the whole region $ \delta\in[-0.5,0.5] $. We see from Fig. \ref{fig: mse vs delta}(a) that the proposed estimator already has a close-to-CRLB performance at some $ \delta $ after the first iteration. We also see that the proposed estimator provides a performance lower bound for GAM. This is expected, since both estimators take $ \delta_0=0.25 $ while the proposed one achieves a more accurate approximation between $ \rho_i $ and $ \xi_i $. We see from Fig. \ref{fig: mse vs delta}(b) that, after the second iteration, the proposed estimator achieves the best flatness in the whole region of $ \delta $.

\begin{figure}[!t]
	\centering{\includegraphics[width=85mm]{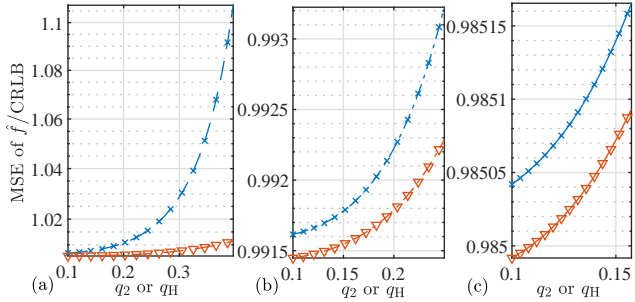}}
	\caption{MSE performance versus $ q_2 $, where $ \gamma=20 $ dB. The sub-figures (a), (b) and (c) are for $ N=8,16 $ and $ 32 $, respectively. For fair comparison with HAQSE, we set $ q_{\mathrm{H}}=q_2 $ which is evenly taken from $ [0.1,N^{-1/3}] $. }
	\label{fig: mse vs q}
	\vspace{-5mm}
\end{figure}

Fig. \ref{fig: mse vs q} compares HAQSE and the proposed estimator in terms of $ q_2 $ (for the proposed) or $ q_{\mathrm{H}} $ (for HAQSE). We see that both estimators show the tight convergence to the CRLB in the case of $ N=16 $ and $ 32 $, and show the robustness against $ q $. This is consistent with the analysis in \cite{FreqEst2019TCOM_Serbes} that the asymptotic performance of HAQSE shall remain the same for $ q_{\mathrm{H}}\le N^{-1/3} $. We also see that our estimators always provides a performance lower bound for the HAQSE across the whole region of $ q_2 $. Notably, we see that, for the small $ N=8 $, our estimator still presents a stable MSE performance over $ q_2 $'s, while HAQSE shows an increasingly worse performance as $ q_{\mathrm{H}} $ increases. The visible improvement achieved by the proposed estimator over the state-of-the-art HAQSE is: \textit{on one hand}, due to the more accurate initial estimate in the first iteration, as has been demonstrated in Figs. \ref{fig: mse vs snr diff N} to \ref{fig: mse vs delta}; and \textit{on the other hand}, ensured by the newly proposed much more accurate approximation between $ \rho_i $ and $ \xi_i $.

\section{Conclusion and Remarks} \label{sec: conclusion}
This correspondence develops an accurate frequency estimator with {fewer} interpolations yet better accuracy and wider applicability to small number of samples, as compared with the state-of-the-art estimators. 
This is achieved by introducing, for the first time, the \PA~in approximating the updating function of estimation error. This is also accomplished by newly unveiled features of the updating function and the derivation of a closed-form solution to solving the residual estimation error in each iteration. 
Extensive simulations validate the performance superiority of the proposed estimator over the state-of-the-art estimators. 

{We remark that the single-tone scenario is focused on in this correspondence for introducing the new ideas. Using the estimate-and-subtract strategy\footnote{{The strategy first estimates each frequency coarsely as a single tone, as if there were no other tones; then, from the second round, each tone is refined by subtracting the recovered signals of other tones. With more rounds of refinement performed, the estimates of all tones can be increasingly accurate.}} (ESS) \cite{FreqEst_AMexMultipleTones2017SP,FreqEst_aoaEstAboutanios2017GRSL,FreqEst_multiToneQSE2020}, a single-tone estimator can often be extended to multi-tone scenarios. 
	For instance, the two benchmarks \AM~\cite{FreqEst_TSP2005AandM} and HAQSE \cite{FreqEst2019TCOM_Serbes}, originally developed for single tone, have been successfully applied to multi-tone scenarios applying ESS \cite{FreqEst_AMexMultipleTones2017SP,FreqEst_aoaEstAboutanios2017GRSL,FreqEst_multiToneQSE2020}. The proposed estimator, combining ESS, is expected to work for multi-tone scenarios as well, since it outperforms \AM~and HAQSE, as presented in Section \ref{sec: simulations}. The detailed extension is deferred to future work.}

\appendix

\subsection{Proof of Lemma \ref{lm: rho monotonic}}\label{app: proof monotonicity and symmetry}
Since $ 0<q_i\le 0.5 $ is satisfied, $ \mathcal{S}(\xi_i,\pm q_i) $ approximates the square of a sinc function \cite{book_oppenheim1999discrete}.
Hence, we know that $ \mathcal{S}(\xi_i,q_i) $ is a monotonically increasing function from $ -q_i $ to $ q_i $ while $ \mathcal{S}(\xi_i,-q_i) $ monotonically decreases in the same region. These can be translated into 
$ \mathcal{S}'_{+}>0,~\mathcal{S}'_{-}<0 $,
where $ \mathcal{S}'_{+} $ denotes the first derivative of $ \mathcal{S}(\xi_i,q_i) $ w.r.t. $ \xi_i $ and $ \mathcal{S}'_{-} $ is similarly defined for $ \mathcal{S}(\xi_i,-q_i) $. Accordingly, the monotonicity of $ f(\xi_i) $ can be deduced from its first derivative, i.e., $ f' 
= {
	2(\mathcal{S}'_{+}\mathcal{S}_{-} - \mathcal{S}'_{-}\mathcal{S}_{+} )
}\big/{[\mathcal{S}_{+}+\mathcal{S}_{-}]^2} >0, $
where $ \mathcal{S}_{+} $ denotes $ \mathcal{S}(\xi_i,q_i) $, and $ \mathcal{S}_{-} $ denotes $ \mathcal{S}(\xi_i,-q_i) $.

From the expression given in (\ref{eq: f( xi_1 )}), we know that $ \mathcal{S}(\xi_i,\pm q_i) $ is symmetric about $ \xi_i=\pm q_i $. Thus, they are also symmetric about the axis of $ \xi_i=0 $ in the region of $ |\xi_i|\le q_i $, i.e.,
\begin{align}
	\mathcal{S}(\xi_i,q_i)= \mathcal{S}(-\xi_i,- q_i),~ \mathcal{S}(-\xi_i,q_i)= \mathcal{S}(\xi_i,- q_i).
\end{align}
Applying this symmetry in (\ref{eq: f( xi_1 )}) leads to
\begin{align}
	f(-\xi_i) &= \frac{ \mathcal{S}(-\xi_i,q_i) - \mathcal{S}(-\xi_i,-q_i) }{ \mathcal{S}(-\xi_i,q_i) + \mathcal{S}(-\xi_i,-q_i) } \nonumber\\
	&= \frac{ \mathcal{S}(\xi_i,-q_i) - \mathcal{S}(\xi_i,q_i) }{ \mathcal{S}(\xi_i,q_i) + \mathcal{S}(\xi_i,-q_i) } = -f(\xi_i),\nonumber
\end{align}
which shows the symmetry of $ f(\xi_i) $ about the origin.

\subsection{Rationale of Setting $ (P,R)=(3,3) $} \label{app: rationale of P=R=3}
{Based on (\ref{eq: tilde f(xi_i) TS}), the TS $ \tilde{f}(\xi_i) $ is of degree six. Thus, according to \cite[Sec. 5.12]{book_press2007numerical}, we have $ P+R=6 $ for the PA $ \hat{f}(\xi_i) $ given in (\ref{eq: hat f(xi_i) PA}). 
	Lemma \ref{lm: rho monotonic} shows $ f(0)=0 $.
	Solving $ \hat{f}(0)=0 $ yields $ a_0=0 $, which then indicates $ P\ge 1 $. Lemma \ref{lm: rho monotonic} also states $ f(-\xi_i)=-f(\xi_i) $. 
	To preserve the odd symmetry, the numerator and denominator of $ \hat{f}(\xi_i) $ can only have odd and even powers of $ \xi_i $, respectively, since there is a non-zero constant $ b_0 $ in the denominator.  
	Based on the above analysis, $ (P,R)=(1,5) $ or $ (2,4) $ leads to the same PA with the degree of the denominator polynomial up to four; $ (P,R)=(3,3) $ or $ (4,2) $ leads to the same PA with the degree of the denominator polynomial up to three; and $ (P,R)=(5,1) $ or $ (6,0) $ makes the PA degenerated into TS. 
	Given that a cubic polynomial can be more tractable than a quartic one, we employ the PA with $ (P,R)=(3,3) $.}

\ifCLASSOPTIONcaptionsoff
\newpage
\fi

\bibliographystyle{IEEEtran}
\bibliography{IEEEabrv,./ref}

\begin{thebibliography}{10}
\providecommand{\url}[1]{#1}
\csname url@samestyle\endcsname
\providecommand{\newblock}{\relax}
\providecommand{\bibinfo}[2]{#2}
\providecommand{\BIBentrySTDinterwordspacing}{\spaceskip=0pt\relax}
\providecommand{\BIBentryALTinterwordstretchfactor}{4}
\providecommand{\BIBentryALTinterwordspacing}{\spaceskip=\fontdimen2\font plus
\BIBentryALTinterwordstretchfactor\fontdimen3\font minus
  \fontdimen4\font\relax}
\providecommand{\BIBforeignlanguage}[2]{{%
\expandafter\ifx\csname l@#1\endcsname\relax
\typeout{** WARNING: IEEEtran.bst: No hyphenation pattern has been}%
\typeout{** loaded for the language `#1'. Using the pattern for}%
\typeout{** the default language instead.}%
\else
\language=\csname l@#1\endcsname
\fi
#2}}
\providecommand{\BIBdecl}{\relax}
\BIBdecl

\bibitem{FreqEst_stallite_wang2019near}
W.~Wang \emph{et~al.}, ``Near optimal timing and frequency offset estimation
  for {5G} integrated {LEO} satellite communication system,'' \emph{IEEE
  Access}, vol.~7, pp. 113\,298--113\,310, 2019.

\bibitem{thesis_freqEst_2002UTS_Aboutanios}
E.~Aboutanios, ``Frequency estimation for low earth orbit satellites,'' Ph.D.
  dissertation, Faculty of Engineering (Telecommunications Group), UTS, 2002.

\bibitem{Kai_freqEstComLttr2020}
K.~{Wu}, W.~{Ni}, J.~{Andrew Zhang}, R.~P. {Liu}, and Y.~{Jay Guo},
  ``Refinement of optimal interpolation factor for {DFT} interpolated frequency
  estimator,'' \emph{IEEE Commun. Lett.}, vol.~24, no.~4, pp. 782--786, 2020.

\bibitem{8057284}
R.~C. {Daniels}, E.~R. {Yeh}, and R.~W. {Heath}, ``Forward collision vehicular
  radar with {IEEE} 802.11: Feasibility demonstration through measurements,''
  \emph{IEEE Trans. Veh. Techn.}, vol.~67, no.~2, pp. 1404--1416, 2018.

\bibitem{Andrew_JCASmultiBeamTVT2019}
J.~A. {Zhang}, X.~{Huang}, Y.~J. {Guo}, J.~{Yuan}, and R.~W. {Heath},
  ``Multibeam for joint communication and radar sensing using steerable analog
  antenna arrays,'' \emph{IEEE Trans. Veh. Techn.}, vol.~68, no.~1, pp.
  671--685, 2019.

\bibitem{FreqEst_TSP2005AandM}
E.~{Aboutanios} and B.~{Mulgrew}, ``Iterative frequency estimation by
  interpolation on {Fourier} coefficients,'' \emph{IEEE Trans. Signal
  Process.}, vol.~53, no.~4, pp. 1237--1242, April 2005.

\bibitem{Freq_liu2011generalizationAM}
Y.~Liu, Z.~Nie, Z.~Zhao, and Q.~H. Liu, ``Generalization of iterative {Fourier}
  interpolation algorithms for single frequency estimation,'' \emph{Digital
  Signal Process.}, vol.~21, no.~1, pp. 141--149, 2011.

\bibitem{FreqEst_SP2014AandM_derivatives}
J.-R. Liao and C.-M. Chen, ``Phase correction of discrete {Fourier} transform
  coefficients to reduce frequency estimation bias of single tone complex
  sinusoid,'' \emph{Signal Process.}, vol.~94, pp. 108 -- 117, 2014.

\bibitem{FreqEst_SPL2017AandM_derivatives}
S.~{Ye}, J.~{Sun}, and E.~{Aboutanios}, ``On the estimation of the parameters
  of a real sinusoid in noise,'' \emph{IEEE Signal Process. Lett.}, vol.~24,
  no.~5, pp. 638--642, May 2017.

\bibitem{FreqEst2019TCOM_Serbes}
A.~{Serbes}, ``Fast and efficient sinusoidal frequency estimation by using the
  {DFT} coefficients,'' \emph{IEEE Trans. Commun.}, vol.~67, no.~3, pp.
  2333--2342, March 2019.

\bibitem{FreqEst_aoaEstAboutanios2017GRSL}
E.~{Aboutanios}, A.~{Hassanien}, M.~G. {Amin}, and A.~M. {Zoubir}, ``Fast
  iterative interpolated beamforming for accurate single-snapshot {DOA}
  estimation,'' \emph{IEEE Geosci. Remote Sens. Lett.}, vol.~14, no.~4, pp.
  574--578, 2017.

\bibitem{Kai_JCAS_fhMIMO_waveform}
K.~{Wu}, J.~{Andrew Zhang}, X.~{Huang}, Y.~{Jay Guo}, and R.~W. {Heath},
  ``Waveform design and accurate channel estimation for frequency-hopping
  {MIMO} radar-based communications,'' \emph{IEEE Trans. Commun.}, pp. 1--1,
  2020.

\bibitem{book_press2007numerical}
W.~H. Press, S.~A. Teukolsky, W.~T. Vetterling, and B.~P. Flannery,
  \emph{Numerical recipes 3rd edition: The art of scientific computing}.\hskip
  1em plus 0.5em minus 0.4em\relax Cambridge university press, 2007.

\bibitem{FreqEst_quinn1994estimating}
B.~G. Quinn, ``Estimating frequency by interpolation using {Fourier}
  coefficients,'' \emph{IEEE trans. Signal Process.}, vol.~42, no.~5, pp.
  1264--1268, 1994.

\bibitem{web_cubicFormulaWolfram}
E.~W. Weisstein, ``Cubic formula,'' \emph{https://mathworld. wolfram. com/},
  2002.

\bibitem{FreqEst_AMexMultipleTones2017SP}
S.~Ye and E.~Aboutanios, ``Rapid accurate frequency estimation of multiple
  resolved exponentials in noise,'' \emph{Signal Process.}, vol. 132, pp.
  29--39, 2017.

\bibitem{FreqEst_multiToneQSE2020}
A.~{Serbes} and K.~{Qaraqe}, ``A fast method for estimating frequencies of
  multiple sinusoidals,'' \emph{IEEE Signal Process. Lett.}, vol.~27, pp.
  386--390, 2020.

\bibitem{book_oppenheim1999discrete}
A.~V. Oppenheim, J.~R. Buck, and R.~W. Schafer, \emph{Discrete-time signal
  processing. Vol. 2}.\hskip 1em plus 0.5em minus 0.4em\relax Upper Saddle
  River, NJ: Prentice Hall, 2001.

\end{thebibliography}

\end{document}